# Unzipping Kinetics of Double-Stranded DNA in a Nanopore

Alexis F. Sauer-Budge[1, 2], Jacqueline A. Nyamwanda[2], David K. Lubensky[3], and Daniel Branton[2]

[1]*Biophysics Program, Harvard University, Cambridge, MA 02138*
[2]*Department of Molecular & Cellular Biology, Harvard University, Cambridge, MA 02138*
[3]*Bell Labs, Lucent Technologies, 700 Mountain Ave., Murray Hill NJ 07974*

**We studied the unzipping kinetics of single molecules of double-stranded DNA by pulling one of their two strands through a narrow protein pore. PCR analysis yielded the first direct proof of DNA unzipping in such a system. The time to unzip each molecule was inferred from the ionic current signature of DNA traversal. The distribution of times to unzip under various experimental conditions fit a simple kinetic model. Using this model, we estimated the enthalpy barriers to unzipping and the effective charge of a nucleotide in the pore, which was considerably smaller than previously assumed.**

Single-molecule techniques allow direct explorations of nucleic acid mechanics, including the stretching and unzipping of double-stranded DNA (dsDNA) [1-3]. Early measurements near equilibrium provide primarily thermodynamic information whereas more recent kinetic approaches have shown that many micromechanical experiments can be understood in terms of one-dimensional energy landscapes along the direction of the applied force [4-7]. Work on a variety of systems [8, 9] has demonstrated that single molecule experiments can reveal behavior that is not detected with ensemble-averaged measurements. Here, we present a new single molecule approach to the kinetics of strand separation in dsDNA. Our approach does not require any covalent modification of the molecules being studied and is well suited to studying strand separation in short oligomers that can be synthesized with any desired sequence. We demonstrate that force-induced unzipping follows a one-dimensional kinetic pathway [5, 6, 9] and use the measured kinetic parameters to infer the effective charge on DNA in the $\alpha$-hemolysin pore [10], a system of interest for its biotechnological applications [11, 12].

To explore strand separation in a nanopore we designed two synthetic DNA constructs, 100/50*comp* and 100/50*mis* (Fig. 1, top), both containing a 50 base pair (bp) duplex region and a 50 base single-stranded overhang. The mismatches in the duplex region of 100/50*mis* made it possible to separately amplify each of the two single-stranded components of the parent molecule using appropriate primers in a polymerase chain reaction (PCR).

Either of these two constructs were added to the receiving, or *cis,* chamber of a device consisting of one protein pore ($\alpha$-hemolysin) in an insulating lipid bilayer membrane separating two solution-filled ($\sim$1M KCl, pH 8.0) compartments [12]. AgCl electrodes, one in each compartment, applied $\sim$120mV bias (*cis* negative). This bias tended to capture and translocate the negatively charged DNA constructs into and through the channel [13]. The voltage bias also induced an ionic current flow that was partially blocked as DNA translocated. The duration of the blockades provided the time measurement for the reported kinetics.

The average blockade duration after either of the two DNA constructs were added to the *cis* chamber was three orders of magnitude longer than with single-stranded DNA (ssDNA) of similar length [12]. To account for these long blockades, we postulated that the overhang (diameter ~1.3 nm) was captured by the pore and rapidly traversed its limiting aperture (~1.5 nm [10]), but that when the double-stranded region of the molecule (~2.5 nm) encountered that aperture, strand translocation slowed drastically or was arrested. We hypothesized that the molecule could then (a) have escaped backwards because of thermal motion, or (b) continued to traverse as dsDNA through a distorted pore, or, more likely, (c) the captured strand could have been pulled through the constriction by the voltage bias as the molecule unzipped. In the last case, the electrostatic force on the DNA is analogous to the mechanical forces used in previous unzipping work [3]. To decide among these alternatives, we determined the 50*mer* and the 100*mer* ssDNA content of the anodic, or *trans*, chamber. If the restricted space in the nanopore caused the dsDNA to unzip, only the strand that had been captured and translocated through the pore should have been detected in the *trans* chamber. Following an experiment using 100/50*mis* in the *cis* chamber, only the 100*mer* strand was seen in the *trans* chamber (Fig. 1, bottom). The fact that a substantial amount of DNA was present in the *trans* chamber rules out alternative (a). Because this DNA consisted of hundreds of 100*mer* strands but no 50*mer* strands, the two strands of 100/50*mis* must have been separated by the translocation process, ruling out (b). Since short blockades consistent with the traversal of detectable levels of contaminating unpaired 100*mer* or 50*mer* from *cis* to *trans* were **not** observed (data not shown), our data indicate that (c) the captured 100*mer* strands of 100/50*mis* had translocated through the constriction without their initial 50*mer* partner. Therefore, unzipping had occurred. Although 50*mer* strands could potentially have traversed the pore after unzipping, our failure to detect them in the *trans* chamber is readily explained by calculating their capture probability, which is related to their *cis* chamber concentration [13, 14] (50mer <$10^{-10}$ M vs. 100/50mis ~1 M).

Next, we looked for the kinetic model with the fewest free parameters that could describe strand separation. If the reaction were a two-state process, with the dsDNA either fully melted or fully paired, we would expect the length of time the polymer blocked the pore to be exponentially distributed. But this time distribution for the 100/50*mis* DNA had a more complicated shape (Fig. 2, top), indicating that the kinetic mechanism included at least one intermediate. We postulated a two-step mechanism (Fig. 3), with step one the reversible unzipping of the DNA up to the 4 bp mismatch, and step two the unzipping of the rest of the strand. Other processes were assumed not to be rate limiting. Using this model, we can solve for the distribution of blockade times $P_{BLOCK}(t)$ assuming that the voltage bias was sufficiently high to maintain $k_{-1} \approx 0$.

$$P_{BLOCK}(t) \approx \frac{k_1 k_2}{k_1 - k_2}\left[e^{-k_2 t} - e^{-k_1 t}\right]. \quad \text{Eq. (1)}$$

Typical results for unzipping experiments were well fit by Eq. 1 for voltages exceeding 130±10 mV, at 1M KCl and 20°C as well as for temperatures exceeding 25±5°C at 1M KCl and 120mV (Fig. 2, top). Similarly straightforward kinetic schemes, with only one or a few intermediates, have been used successfully to describe other single-molecule force-induced strand separation experiments [9, 15].

If the intermediate was the dsDNA unzipped to the four-base mismatch, then a molecule without those mismatches should not exhibit the intermediate state, and its kinetics should be that of a first order reaction, with a longer mean blockade time than for 100/50*mis*. As predicted, the distribution of blockade durations for 100/50*comp* was exponential and the mean time was more than two times longer than for 100/50*mis* (Fig. 2, bottom). The ability to observe the altered DNA unzipping kinetics after inserting or removing mismatches underscores the power of this method. In general, there should be kinetic intermediates wherever there is a sufficiently pronounced minimum in the unzipping energy landscape [5], with mismatches giving rise to especially deep energy wells. It should thus be possible to relate the kinetics of more complicated reactions, with multiple intermediates, to the dsDNA sequence [16].

We explored the parameter space covered by the model by varying the voltage applied, the temperature, and the buffer ionic strength. For all tested conditions, the rate constant for step one of 100/50mis unzipping was considerably larger than for step two (Fig. 4). This observation presumably reflects the presence of a larger free energy barrier for the second step, perhaps related to the destabilizing effect of the four-base mismatch on the first, shorter duplex section that is proximal to the overhang.

We expected the force on the single strand, and thus the electrostatic energy gained from translocating each additional nucleotide, to scale linearly with voltage. As long as the ssDNA completely threads the pore and the voltage drop falls mostly across the neck of the channel (Fig. 3), the change in electrostatic energy when the polymer moves forward one nucleotide is independent of both the length of the pore and the number of nucleotides already translocated [17]. If the transition state of a given unzipping reaction occurs after $n_{max}$ bases have been unzipped, its energy should thus depend on voltage as $-qn_{max}V$, where V is the applied voltage and q can be interpreted as the effective charge on a nucleotide in the pore. This implies

Eq. (2)

$$k \approx A e^{-\left(\frac{\Delta G^*}{k_B T}\right)}$$
$$\Delta G^* \approx \Delta G^{*o} - qn_{max}V$$

where k is the rate constant, $\Delta G^*$ is the energy barrier with voltage applied, $\Delta G^*_0$ is the energy barrier to the reaction at V=0, and A is an unknown constant [4, 18].

The backbone of ssDNA has one negatively charged phosphate per nucleotide. If, as expected, counterions are transported across the membrane along with the DNA, then the electrostatic energy gained per translocated nucleotide is smaller than in a salt-free environment. The effective charge defined by Eq. (2) is then less than one full electron charge (e) per nucleotide. The precise value of *q* could be affected by the polar groups on the walls of the protein pore, the extended conformation of the DNA strand, and the altered ion flow caused by the constriction. Thus the value of *q* within the pore could differ substantially from that in free solution [19]. By fitting the plots of rate constants vs. voltage and inferring $n_{max}$ from computationally generated energy landscapes [5, 20], we estimated the effective charge on a nucleotide traversing the pore to be $q \approx 0.1e$. Although this value might be in error by as much as a few tenths of an electron charge because of the simplifications in our model and limited data, the same estimate for *q* was obtained from the voltage dependence of the rate constant for unzipping 100/50*comp* at 1M KCl (data not shown), and from those describing 100/50*mis* unzipping at both 1M and 1.3M KCl (Fig. 4).

While holding voltage and ionic strength constant, we raised the temperature to measure the barriers to unzipping. The rate constants depended exponentially on $1/k_B T$ (Fig. 4). Using Eq. 2 with $\Delta G^* = \Delta H^* - T\Delta S^*$, we calculated the enthalpy barrier $\Delta H^*$ for each step in the unzipping reaction. As one might predict from the fact that $k_2 < k_1$, $\Delta H^*$ for the second step was 35-70 kJ/mol higher than for the first step and depended on the ionic strength of the buffer (Table 1). Note that as with conventional DNA melting [21], a large entropic gain upon denaturation is expected to make the $\Delta G^*$ values substantially smaller than our measured $\Delta H^*$'s.

Indeed, our fitted $\Delta H^*$ values were of the same order of magnitude as measured thermodynamic parameters for the melting of $n_{max}$ bps in free solution [21]. Because the environment in the pore differs from that used for melting measurements, we would not expect any closer agreement.

The stabilizing effect of salt on the DNA backbone has been evaluated thermodynamically by overstretching [1] and thermal melting [22]. We explored the stability of the 100/50*mis* at several salt concentrations, two of which are presented (Fig. 4). As expected, the rate constants consistently decreased with increasing KCl. The heat capacity of DNA accounts for the stronger stabilization of A-T bps relative to G-C bps with increasing ionic strength [23]. Therefore the fact that $\Delta H^*$ increased more with increasing salt concentration for the second step of 100/50*mis* unzipping than it did for the first step can be at least partially explained by the higher fraction of A-T bps in the second portion of the molecule (57% vs. 43%).

In contrast to many single molecule experiments, our measurements of unzipping kinetics in a nanopore enable one to explore the distribution of molecular properties for many individual molecules in an efficient manner and without labeling or covalent modifications [8, 9, 15]. The mechanism of unzipping, whose details are governed by the environment that the DNA experiences (ionic strength, temperature, voltage) and by the DNA sequence, can be deduced directly from the unzipping time distribution. Moreover, by examining the unzipping behavior of dsDNA while varying voltage and temperature, one can measure several biophysical properties of the molecules, including the unzipping reaction rate constants, the effective charge on the nucleotide as it passes through the protein pore, and the enthalpy barrier height for the reaction. Our inferred value of the effective charge $q \approx 0.1e$, while providing only a rough estimate, sheds light on the behavior of screening in the confined environments typical of biological systems and improves our basic understanding of a system, the α-hemolysin pore, that has been the focus of substantial biotechnological activity. Although the channel we used allows some of the parameter space to be investigated, it will be advantageous to access more stringent conditions than are tolerated by a protein pore. Advances in fabrication techniques should soon make it possible to replace the protein nanopore with a solid state pore fabricated to the desired size and able to withstand a broad range of solvents and temperatures [24]. With such a nanopore, the rapid single molecule method described here could investigate interactions as diverse as inter-strand base pairing, the binding of transcription factors to dsDNA, and the movement of enzymes along the length of a polymer.

Acknowledgements: We thank Drs. M. Muthukumar, D. Deamer, and M. Akeson for their helpful comments, E. Brandin for technical assistance with the PCR experiments, and Dr. D. Nelson for related collaborations. This work was supported by DARPA and the US DOE.

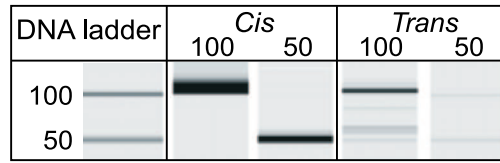

FIG. 1. Top: The sequence of 100/50*mis*, which consisted of a 50 bp region (shown) plus a 50 base ssDNA 3'-overhang (the first 5 bases are shown). The sequence of 100/*50comp* was identical except that the mispaired bases in the *50mer* strand (underlined) were replaced by bases complementary to the 100*mer* strand. Bottom: Gel view of a capillary electrophoresis analysis of PCR amplified content of the *cis* and *trans* chambers after unzipping 100/50*mis* through the nanopore.

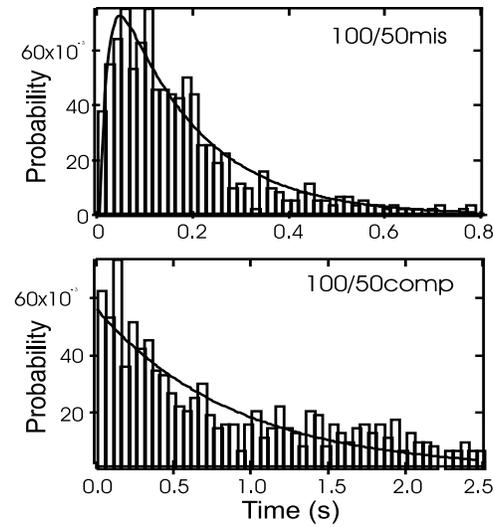

FIG. 2 The distribution of blockade durations for 100/50*mis* (top, mean time is 185 ms), and 100/50*comp* (bottom, mean time is 435 ms) at 1M KCl, 20°C, and 140mV. Only those events that block the pore to the same degree as 100*mer* control ssDNA are considered. The solid lines are fits to the kinetic models discussed in the text.

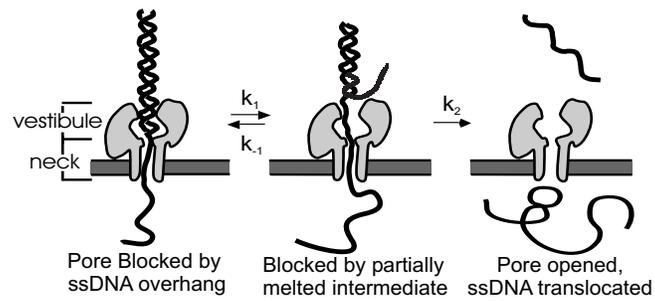

FIG. 3 Proposed mechanism of 100/50*mis* unzipping during translocation. We assume that most of the voltage drop occurs in the transmembrane neck of the channel, with little voltage change across the wide vestibule [25].

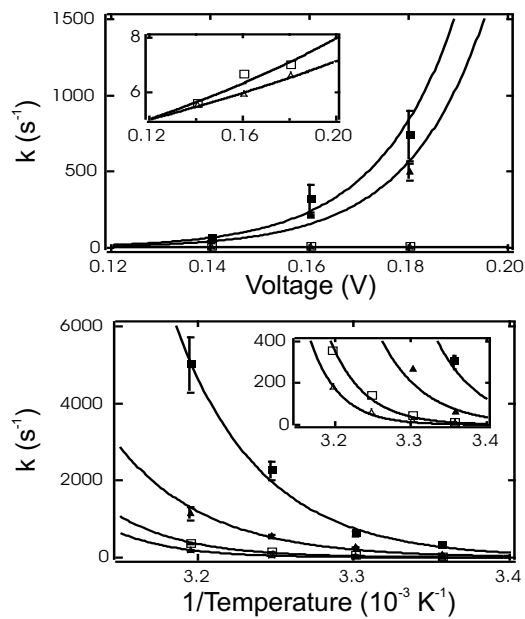

FIG. 4. The effect of voltage (top, at 20°C) and temperature (bottom, at 120 mV) on the unzipping rate constants of 100/50*mis*. Filled symbols, $k_1$; open symbols, $k_2$; squares, 1M KCl, triangles 1.3M KCl. The insets show expanded y-axes. The lines are fits to Eq. 2. The error bars reflect the uncertainties in the fitted rates due to counting error in our time histograms.

Table I: The calculated enthalpy barriers for the kinetic steps of unzipping 100/50*mis* at 120 mV applied voltage.

| [KCl] (M) | Step in reaction | $\Delta H^*$ (kJ/mol) |
|---|---|---|
| 1 | 1 | 146±13 |
| 1.3 | 1 | 149±29 |
| 1 | 2 | 184.5±1.0 |
| 1.3 | 2 | 222.2±2.6 |

[1] H. Clausen-Schaumann, et al., Biophys. J. **78**, 1997 (2000); J. R. Wenner, et al., *ibid.* **82**, 3160 (2002).
[2] I. Rouzina and V. A. Bloomfield, Biophys. J. **80**, 882 (2001); **80**, 894 (2001); T. R. Strick, et al., Prog. Biophys. Mol. Biol. **74**, 115 (2000); A. Noy, et al., Chem. Biol. **4**, 519 (1997); T. Strunz, et al., Proc. Natl. Acad. Sci. U. S. A. **96**, 11277 (1999).
[3] M. Rief, H. Clausen-Schaumann and H. E. Gaub, Nature Struct. Biol. **6**, 346 (1999); U. Bockelmann, et al., Biophys. J. **82**, 1537 (2002).
[4] E. Evans, Annu. Rev. Biophys. Biomol. Struct. **30**, 105 (2001).
[5] D. K. Lubensky and D. R. Nelson, Phys. Rev. E **65**, 31917 (2002).
[6] S. Cocco, R. Monasson and J. F. Marko, Proc. Natl. Acad. Sci. U. S. A. **98**, 8608 (2001).
[7] B. Isralewitz, M. Gao and K. Schulten, Curr. Opin. Struct. Biol. **11**, 224 (2001).
[8] X. S. Xie, Single Molecule **2**, 229 (2001); X. Zhuang, et al., Science **288**, 2048 (2000).
[9] J. Liphardt, et al., Science **292**, 733 (2001).
[10] L. Song, et al., Science **274**, 1859 (1996).
[11] W. Vercoutere, et al., Nature Biotech. **19**, 248 (2001); S. Howorka, S. Cheley and H. Bayley, *ibid.* **19**, 636 (2001).
[12] J. J. Kasianowicz, et al., Proc. Natl. Acad. Sci. U. S. A. **93**, 13770 (1996); M. Akeson, et al., Biophys. J. **77**, 3227 (1999); A. Meller, et al., Proc. Natl. Acad. Sci. U. S. A. **97**, 1079 (2000).
[13] A. Meller and D. Branton, Electrophoresis **23**, 2583 (2002).
[14] S. E. Henrickson, et al., Phys. Rev. Lett. **85**, 3057 (2000).
[15] R. Merkel, et al., Nature (London) **397**, 50 (1999).
[16] S. Cocco, J. Marko and R. Monasson, cond-mat/0207609; U. Gerland, R. Bundschuh and T. Hwa, cond-mat/0208202.
[17] D. K. Lubensky and D. R. Nelson, Biophys. J. **77**, 1824 (1999).
[18] G. Bell, Science **200**, 618 (1978).
[19] G. S. Manning, Q. Rev. Biophys. **11**, 179 (1978).
[20] If the free energy gained from unzipping is more than the cost of breaking a given bp, then that bp will contribute a downward step to the energy landscape. If this occurs, $n_{max}$, the location of the highest energy barrier along the unzipping pathway, need not refer to the last bp before a mismatched region [5,17].
[21] K. J. Breslauer, et al., Proc. Natl. Acad. Sci. U. S. A. **83**, 3746 (1986).
[22] C. Schildkraut and S. Lifson, Biopolymers **3**, 195 (1965).
[23] I. Rouzina and V. A. Bloomfield, Biophys. J. **77**, 3242 (1999).
[24] J. Li, et al., Nature (London) **412**, 166 (2001).
[25] A. Meller, L. Nivon and D. Branton, Phys. Rev. Lett. **86**, 3435 (2001).